\documentstyle[12pt,twoside]{article}
\begin{document}
\title{The Energetics and Electronic Structure of \\
Defective and Irregular Surfaces on MgO}
\author{L. N. Kantorovich\thanks{On leave from University of Latvia,
19 Rainis, Riga, LV-1050, Latvia}, J. M. Holender and M. J. Gillan \\
Physics Department, Keele University \\
Keele, Staffordshire ST5 5BG, U.K.}
\maketitle
\begin{abstract}
{\em Ab initio} calculations based on the density-functional
pseudopotential approach have been used to study the fully
relaxed structure, the electron distribution and the electronic
density of states of (001) terraces, steps, corners and reverse
corners, and of F-centers at these surface features on MgO. The
calculations confirm earlier predictions of the relaxed structures of
surface irregularities based on simple interaction models. A substantial
narrowing of the band-gap is found at the surface, which for terraces
and steps is due to surface states at the bottom of the conduction
band, but for the corner and reverse corner is also due to surface
states at the top of the valence band. The F-center formation energy
decreases steadily as the coordination of the oxygen site is
reduced. The energy of the F-center level shows a tendency to
approach the top of the valence band as the coordination of
its site decreases.
\end{abstract}

\section{Introduction}

There is a general consensus that surface defects such as
steps, kinks, corners and surface anion and cation vacancies
play a crucial role in molecular processes at oxide surfaces \cite{rew,
k11,k10,k5,k7,k8,k9}.
Yet in a very general sense the background understanding needed
to discuss these processes is seriously lacking. There is
very little direct experimental information about the structure and
energetics of these surface defects for most oxides (see, for instance,
references \cite{k11,k5}). It is true
that there has been useful theoretical work on the relaxed structure
of steps and other surface irregularities (see e.g. references
\cite{k9,k4,k6})
based on simple (and usually
empirical) interaction models, but the validity of the assumptions
on which the models are based is not always clear. Important
insights have also been gained {\em via} semi-empirical quantum
calculations \cite{k10,k2,k1,ds2}.
In a few cases, fully {\em ab initio} calculations
on oxide surfaces have been reported \cite{k3,s13,ds10,s5,s12,s4,har},
but these are almost
all restricted to flat non-defective surfaces
(see, however, Ref. \cite{k3,har}).

The aim of the present paper is to contribute to the background
understanding of defective oxide surfaces by reporting a series of
{\em ab initio} total-energy calculations on the relaxed atomic
structure and the electronic structure of the flat MgO (001)
surface, of steps, corners and reverse corners (see Section 5 for the
complete explanation of these terms) on this surface,
and of surface F-centers (two electrons trapped in an oxygen
vacancy) at the flat surface and at the
step, corner and reverse corner.  For reference purposes, we also present
results on the bulk F-center. Our particular interest is in the
processes of oxygen exchange between the gas phase and the crystal lattice,
and the present work is intended to provide some of the essential
information needed for describing these processes, but we
believe our results will also be relevant to other problems, such
as the interpretation of the spectroscopic properties of the MgO surface.

Our calculations are based on the density-functional pseudopotential
approach \cite{striv,gillan,payne}
which has already proved highly successful in many
previous studies of oxide surfaces \cite{ds10,man-gil,man-gil1}
and point defects in oxides \cite{vita,db2}. In order
to treat defective and irregular oxide surfaces, one needs to
do large numbers of calculations on systems containing up to ca. 60
atoms with full relaxation of the system to mechanical equilibrium.
With the {\em ab initio} techniques used in the present work, this
can now be done fairly routinely.

The paper is organized as follows. In the following section, we summarize
the techniques we have used for calculating the ground-state total
energy of the system and the electronic density of states. We then present
our results on the bulk F-center (Section 3). This is followed by a
description of our results for the flat (001) surface with and without
F-centers (Section 4) and for the step, corner and reverse corner and for
F-centers at these features (Section 5). A discussion of our results
is given in Section 6.

\section{Methods}

\subsection{Total energy calculation}

Density-functional theory and the pseudopotential approximation
are well established and widely used methods, which have been extensively
reviewed in the literature (see e.g. references \cite{striv,gillan,payne}).
Only valence electrons
are represented explicitly in the calculations, the valence-core
interaction being described by non-local pseudopotentials which
are generated by {\it ab initio} calculations on isolated atoms. The
effects of electron correlation are included using the local density
approximation (LDA).
A large amount of work on MgO based on this approach has already been reported
\cite{s13,ds10,s4,db2,b2,db3,db4,db18,db22,b9}.
The calculations are
performed on periodically repeated cells, with every occupied valence
orbital represented as a plane-wave expansion. This expansion includes
all plane waves whose kinetic energy $E_{k} = \hbar^{2}k^{2}/2m$
({\bf k} the wavevector, {\it m} the electronic mass) satisfies
$E_{k} < E_{cut}$, where $E_{cut}$ is a chosen plane-wave cut-off energy.
This means that calculations can be taken to convergence with respect
to the size of basis set simply by increasing the cut-off.

In the present work, the self-consistent ground state of the system
is determined using the Car-Parrinello approach \cite{car-par},
in which the total energy
is minimized with respect to the plane wave coefficients of the
occupied orbitals, the minimization being performed by the
conjugate-gradient technique \cite{payne}.
The calculations were performed with the code CETEP
running on the Cray T3D parallel supercomputer at the Edinburgh Parallel
Computer Centre. The computational strategy underlying the CETEP code
has been described in the literature \cite{cetep},
but the code has been extensively
rewritten to take full advantage of the T3D. The main change concerns
the way the conjugate gradient search is done. In previous versions,
each band was updated sequentially in the iterative search for the ground
state. In the version used here, all occupied bands are updated
simultaneously. This method exploits the large memory available
on the T3D, and results in a speed-up of about a factor of three
for the present type of problem.

Technical details of the calculations are as follows. The pseudopotentials
for Mg and O atoms are identical to those used in our recent calculations
on the adsorption of NH$_{3}$ at the MgO (001) surface \cite{ds10}.
As in that work,
the Kleinman-Bylander representation \cite{ps} of the pseudopotentials is
used, with the {\it s} component treated as local for Mg and the {\it d}
component for O.
The only difference with this work lies in using the real-space representation
\cite{rps} of the pseudopotentials, which speeds up the calculations
significantly for systems of the large size treated here.
We demonstrated  in our earlier work \cite{ds10} that with a plane-wave
cut-off $E_{cut}$ = 600 eV the total energy is essentially perfectly
converged with respect to the size of the basis set, and the same
cut-off has been used here. Electronic exchange and correlation are
represented by the commonly used Ceperley-Alder form \cite{exch1,exch2}.
Brillouin zone sampling for the ground state calculations is performed
using the Monkhorst-Pack scheme \cite{mp}, as noted in more detail later.
It was found in our previous calculations \cite{ds10} that in equilibrium
the nearest-neighbor distance is $d=2.082$ \AA, and we have used the same
value in our present calculations. Note that it
is close to the experimental value 2.105 \AA\ \cite{wyckoff}.

\subsection{Electronic density of states (DOS)}

For all systems studied in this paper we have calculated the electronic
density of states (DOS)
$$
{\cal N}(\epsilon) = {1\over N}
\sum_{n{\bf k}} \delta \left( \epsilon -\epsilon _{n{\bf k}} \right) ,
\eqno (1)
$$
where $\epsilon _{n{\bf k}}$ is the
eigenvalue of the Kohn-Sham equation for the
band $n$ and the wave vector {\bf k} in the Brillouin
zone (BZ) of the supercell,
{\it N} is
the number of unit cells in the crystal. When making
calculations on {\it bulk} systems,
the standard tetrahedron  method has been used \cite{tetr1,tetr2}.
First, we note
that the BZ can be chosen as a parallelepiped defined by the
three primitive reciprocal lattice vectors ${\bf B}_{1}$, ${\bf B}_{2}$,
${\bf B}_{3}$. Then, we divide the BZ into small parallelepipeds of an
equal volume by dividing every lattice vector ${\bf B}_{i}$ ($i=$1,2,3)
by integers $n_{1}$, $n_{2}$, $n_{3}$. Every small parallelepiped
is then split into six tetrahedra of an equal volume \cite{tetr3},
and all the
{\bf k}-points at the edges of the tetrahedra form a submesh of
{\bf k}-points. Using the point-group symmetry operations of the system,
a significantly reduced set $\Pi$
of symmetry inequivalent {\bf k}-points is obtained, and an array
containing the correspondence between these {\bf k}-points and all
edges of all the tetrahedra is generated. Then,
using the CASTEP code \cite{payne},
we calculate the eigenvalues
$\epsilon _{n{\bf k}}$ for every {\bf k}-point belonging to the set $\Pi$.
Using the array which sets up the
correspondence between the set $\Pi$ and all edges
of tetrahedra, the total DOS is finally calculated.

Several points are worth mentioning. First, we emphasize that
for systems of the size treated here the calculations of the
eigenvalues $\epsilon _{n{\bf k}}$ are extremely demanding and we
had to restrict our {\bf k}-point sampling by choosing equal
divisions $n_{1} = n_{2}=n_{3}=5$ corresponding to 750 tetrahedra
in the whole BZ. For high-symmetry systems without an F-center (except
for the perfect corner - reverse corner system which has no symmetry at all)
we have used also a larger number of divisions to study the convergence
of the total DOS. We found this division to be reasonable and reliable
as far as the boundaries and the general structure of bands are concerned.
However,
we cannot rely on the fine structure of the total DOS since the energy
resolution appears to be not completely satisfactory.
Second, when making DOS calculations for the surface systems in slab
geometry, it is unnecessary to consider the
{\bf k}-point sampling along the direction
of the normal to the surface \cite{ds10},
so that the BZ is reduced to the two-dimensional
{\it surface} BZ. In this case,
the method of tetrahedron
is simplified to the triangle method.
Third, since the
{\bf k}-point sampling was not very extensive in all our calculations,
we found that often there is an accidental equivalence of some edges of
one and the same tetrahedron which results in an accidental degeneracy
of the eigenvalues, $\epsilon _{n{\bf k}}$. This forced us to modify
slightly the standard method of tetrahedra to cope with this
problem. Namely, we have formally removed this degeneracy by adding
different $\delta_{\alpha} >0$ to the eigenvalues belonging to the
equivalent edges. Then, using equations of the standard method,
the corresponding
modified formulae have been worked out by making the limit
$\delta_{\alpha} \rightarrow +0$.

\section{The bulk F-center}

We have used the techniques described in the previous section
to perform calculations on the bulk F-center using repeating
cells containing 8, 16, 32, 54 and 64 lattice sites. These are
all the repeating cells containing between 8 and 64 sites, which
preserve the O$_{h}$ symmetry of the F-center, and they give
periodic arrays of F-centers having s.c., f.c.c., b.c.c., f.c.c.
and s.c. translational symmetry, respectively. The next available
size of the cell would contain 128 sites. For each repeating cell
we have first calculated the total energy of the perfect crystal.
We have then removed an oxygen atom and repeated the calculation.
In addition to treating the unrelaxed F-center, we have also
studied its equilibrium by relaxing all ions in the system. All
calculations have been performed using the lowest-order
Monkhorst-Pack set of {\bf k}-points, consisting of the eight points
$(\pm {1\over 4}, \pm {1\over 4}, \pm {1\over 4})$ (in units of the primitive
reciprocal lattice vectors). As
usual, it is unnecessary to include ${\bf k}$-points related by inversion
symmetry, so that the calculations are actually performed with
the four points $({1\over 4}, \pm {1\over 4}, \pm {1\over 4})$.

We first report our results for the oxygen removal energy.
This is defined to be the energy (per cell) of the system
containing the F-center, plus the energy of an isolated oxygen
atom minus the energy of the perfect crystal calculated with
exactly the same cell. In view of the large energies involved, it
is important that the same cell be used for the F-center and
perfect crystal to ensure cancellation of errors. The energy of
the isolated oxygen atom was obtained by calculations on a number
of periodic systems containing one oxygen atom per cell. The
final size of the repeating cell in these calculations has been
taken large enough to ensure that the interaction between oxygen
atoms introduces an error of less than 0.01 eV. Since an oxygen
atom in the ground state has a multiplet structure $^3$P$_2$ \cite{g1},
and
the state calculated by the LDA method belongs to the excited state
$^1$D of the atom,
a correction has to be made for the energy of the oxygen atom.
As an estimate for this correction, we have used the experimental
energy difference for these states \cite{g1} which is 1.967 eV.
Note that the local spin density (LSD) calculations give a somewhat
similar correction (the difference between energies calculated
using LSD and LD methods) which is 1.496 eV \cite{b2,db3} and
1.4 eV \cite{b5}.

To estimate the overall error in our oxygen removal energies, we have also
calculated the cohesive energy
(per formula unit with respect to free neutral atoms)
of the MgO crystal. Using the total energy
of the largest repeating cell for the
bulk containing 64 sites and the calculated
total energies of the O and Mg atoms, we have calculated
the cohesive energy to be 11.14 eV.
Note that we have also used the 0.14 eV correction \cite{b2} for the
zero-point vibrational energy of MgO.
Our value turns out to overestimate the experimental cohesive energy
(10.3 eV \cite{cohexp}).
The existence of this discrepancy with experiment of 0.8 eV suggests
that there may be a similar
systematic error in calculated oxygen removal energies.

Our calculated oxygen removal energies for both the unrelaxed
and relaxed F-center are reported in Table 1. Two main
conclusions are clear from these results. First, the removal
energies are completely converged with respect to the size of the
cell already for the 16-site system. In fact, even the 8-site
result differs from the others by only 0.1 eV. Second, the effect
of relaxation is completely negligible. The almost complete
absence of relaxation effects for the neutral F-center in MgO has
also been demonstrated by earlier calculations \cite{db3,db18,db12,db13}.
The comparison
of our calculated oxygen removal energy with experiment and with
other calculations will be discussed later.

Although the relaxations are extremely small, they do show
systematic behaviour for all the sizes of cell we have studied.
The Mg ions adjacent to the F-center relax  outwards by an amount
which is 0.01 \AA\ for the 54-site and 64-site cells and is somewhat
smaller for the smaller cells. The oxygens nearest to the F-center
relax outwards by about 0.005 \AA\ for the two largest cells.
The relaxations surrounding the F-center in MgO have been studied
previously by {\it ab initio} calculations \cite{db3,db18,db12,db13,db5}.
All these calculations agree that the displacements are extremely
small and of the same general magnitude that we have found, but
there are some disagreements about the directions of the
displacements. These disagreements cannot be considered very
significant, given the very small effects involved.

We have calculated the electronic DOS for
the perfect crystal and for the system containing an F-center,
using the tetrahedron technique described in the previous
Section. It is straightforward to use enough ${\bf k}$-points to obtain
good energy resolution for the perfect crystal, but the
calculations on large cells are much more demanding, and the
energy resolution for the F-center systems is less good.
We show in Fig. 1 our results for the DOS of the perfect
crystal and for the 16-site and 54-site F-center systems.

As is well known from many previous calculations, the
valence DOS of the perfect crystal consists of a rather narrow band
of O(2{\it s}) states, and
a broader two-peak band of O(2{\it p}) states.
We have found in our calculations that the widths of these bands
are 2.1 eV and 4.9 eV, respectively, and we estimate the separation
between two peaks in the band of O(2{\it p}) states to be ca. 3.2 eV.
Our calculated separation between bands of O(2{\it s}) and O(2{\it p})
states is ca. 11.3 eV.
The widths, positions and shapes
of the valence bands in the DOS are in close agreement with previous
theoretical results \cite{s13,b2,b9,b5,b7}.
Agreement with experiment
\cite{b10,gap1,gap2,gap3}
(2.0 - 2.5 eV and 5.0 - 7.0 eV for the widths of the two
bands and 13 - 14.5 eV for the {\it s-p}-bands separation) is somewhat worse
but this is usual with LDA calculations. In addition, the bulk
band gap between the top of the upper valence band and the bottom of the
conduction band is underestimated. Our calculated value is 4.8 eV, which
should be compared with the experimental value of 7.8 \cite{gap} (note
that a bigger value 8.7 eV is reported in \cite{k5}).
This degree
of underestimation of band gaps is very typical for LDA
calculations.

The states associated with the F-center are clearly visible in
the lower half of the band gap in Fig. 1.
An isolated F-center would have a single sharply
defined level. In the periodic systems treated here, this level
is broadened into a band. As expected, this band is broader for
the smaller system. The width of this band is 1.6 eV and 0.6 eV
for the 16-site and 54-site systems. The large width for
the small system suggests that the F-center wavefunction is
rather extended, and we shall analyze this question below.
The average energy (i.e.
the first moment) of the F-center band is ca. 2.5 eV above the valence
band maximum (VBM) for the 16-site system and 2.7 eV above the
VBM for the 54-site system.
This can be compared with the position deduced from optical absorption
measurements.  The observed F-center absorption energy is ca. 5.0 eV
\cite{add,db24,db15,db9} and it is generally agreed \cite{db24,db9,db10}
that this represents a
transition from the ground state to a state lying just below
the conduction band. Taking the band gap to be 7.8 eV \cite{gap},
we deduce that the F-center level should be roughly 2.8 eV above the VBM.
Our calculated position
of the F-center level is also in close agreement with
values obtained in other {\it ab initio}  calculations (2.34 eV \cite{db3},
2.55 eV \cite{db4}).

A contour plot of the valence electron density in the 54-site
F-center system is shown in Fig. 2. As expected from earlier work \cite{db3},
the density has a maximum at the center of the vacancy, but the
electron distribution in the vacancy region is rather broad.
To understand this in more detail, it is useful to compare the
electron distributions in the F-center system and the perfect
crystal. We have done this by calculating for each system the
total valence charge inside spheres of different radii $R$
centered at the vacancy. We denote this total charge by $n_{F}(R)$
for the F-center system and $n_{P}(R)$ for the perfect crystal. We
are interested in finding out how rapidly the disturbance due to
the F-center dies away with increasing radius. We therefore
subtract $n_{P}(R)$ from $n_{F}(R)$. To allow for the fact that there
are six more electrons in the perfect crystal, it is convenient
to study the quantity
$$
\Delta n(R) = n_{F}(R) - n_{P}(R) + 6 .
\eqno (2)
$$
With this definition, $\Delta n(R)$ should tend to zero for large
$R$. Our results for $\Delta n(R)$ are shown in Fig. 3. These results
show that the electron distribution outside a radius of
approximately 2.0 \AA\ is essentially the same in the F-center
system as in the perfect crystal. Bearing in mind that the
distance to the nearest Mg ion is 2.08 \AA, this rather good
localization of the charge associated with the F-center explains
the almost complete absence of relaxation effects.

We have also investigated the electron density associated
with the F-center band itself. A contour plot of this density is
shown in Fig. 4. It is clear from this plot that the density of
the F-center band is not completely localized in the vacancy, but
has considerable weight on the nearest and even next-nearest
oxygen ions. We have analyzed this quantitatively and found that
there is roughly one electron in the vacancy, 0.3 electron on
nearest oxygens and the remaining 0.7 electron is distributed in
weak features in a more distant regions. This indicates that the
electron density on oxygen neighbors is associated with states
both at the valence band energies and also at
the energy of the F-center level. Conversely, the electron
density in the vacancy is associated with states both at the
energy of the F-center level and also at the valence-band
energies. These results are in accord with the calculations of
Wang and Holzwarth \cite{db3}.

\section{F-center at the flat (001) surface}

\subsection{The perfect surface}

Before describing our calculations on the surface F-center,
we present briefly our results for the perfect surface. In
all our surface calculations, we use slab geometry, so that the
system we study is a periodically repeated stack of slabs
separated by vacuum layers. For the calculation on the perfect
surface, we use a vacuum width of 4.164 \AA, which our previous
work \cite{ds10} on the MgO (001) surface showed to be enough to make the
interaction between slabs negligible. In the present
calculations, we have studied slabs containing 2, 4 and 6 layers
with 16, 32 and 48 ions respectively in the surface unit cell,
and we found that the
properties of interest are essentially the same in all three
cases. We have performed ${\bf k}$-point sampling using the lowest-order
Monkhorst-Pack scheme. Since sampling along the surface normal is
unnecessary \cite{ds10},
this means that we use the two {\bf k}-points $({1\over 4},\pm {1\over 4},0)$
where the third component is along the surface normal.

There have been a number of experimental studies \cite{s6,s2,welton} and
previous calculations  \cite{k9,ds10,s5,s8,sjacek}
on the properties of the (001) surface,
which indicate that
relaxation effects are extremely small. This is fully confirmed
in the present calculations. In accord with all previous work, we
find that surface oxygens relax outwards and surface magnesiums
relax inwards. For the six-layer slab, the values of these
displacements are $\epsilon _{\rm O}= 0.022$ \AA\ and
$\epsilon _{\rm Mg}= - 0.018$ \AA. These
values are in qualitative agreement with those obtained in our
recent pseudopotential LDA calculations \cite{ds10} using a three-layer
slab, which gave $\epsilon _{\rm O}= 0.032$ \AA\ and
$\epsilon _{\rm Mg}= - 0.003$ \AA.

The total electronic DOS for the unrelaxed six-layer slab
is shown on the top panel of Fig. 5. The most striking feature is the marked
reduction of the band gap. This has a value of 3.2 eV, which is
1.6 eV smaller than our calculated value for the perfect crystal.
The figure indicates that the gap states are pulled down from the
conduction band, rather than splitting from the top of the
valence band.

Our results concerning the reduction of the band gap as one goes
from the bulk to the surface are in good agreement with the measurements
by  energy-dependent
electron-energy-loss spectroscopy (EELS) \cite{k11,s14} where a 2 eV
reduction of the band gap has been observed. A reduction of the
band gap has been also obtained in a number of
previous calculations \cite{k10,s13,sjacek,ds1} though the calculated
values differ.
For example,
surface states below the bottom of the conduction
states were found in recent {\it ab initio} calculations \cite{s13},
which gave the reduction of the band gap by ca. 0.6 eV.
We shall return to the discussion about the surface states in Section 6.

In some oxides, it has been found \cite{2s-split} that the change of the
Madelung potential at the surface causes the O(2{\it s}) states of
surface oxygens to split off from the top of the O(2{\it s}) band. We
do not observe this in the present case. This is because the
Madelung potential is almost unchanged at surface oxygen sites
(the Madelung constant at a (001) surface site in the rock-salt structure
is 1.682 \cite{madsurf}, while the value at a bulk site is 1.748)
and that is consistent with the absence of oxygen-based surface
states.

We have investigated the nature of the surface states by
calculating the fictitious charge density associated with several
of the lowest unoccupied bands. In practice, we have calculated
the quantity $\rho _{i}({\bf r})$ defined by
$$
\rho _{i}({\bf r}) = \sum^{}_{{\bf k}} w_{{\bf k}}\left|\begin{array}{c}
\psi _{i{\bf k}}({\bf r})\end{array}\right| ^{2},
\eqno (3)
$$
where $\psi _{i{\bf k}}({\bf r})$ is the eigenfunction
of band $i$ at wavevector
${\bf k}$ from the irreducible wedge of the BZ,
and $w_{{\bf k}}$ is the weighting factor for this wavevector. The
quantity $\rho _{i}({\bf r})$ represents the electron density that would be
associated with orbital $i$ if the orbital were occupied. Fig. 6
shows $\rho _{i}({\bf r})$ for the lowest unoccupied band, i.e. the band
immediately above the top of the gap. The localization of $\rho _{i}({\bf r})$
in the surface region confirms that this band consists of surface
states. In contrast to the results of another {\it ab initio} LDA
calculation \cite{s13} where surface states were found to be
predominantly of Mg {\it s} character,
our surface states are concentrated mainly  on surface oxygens and above
surface magnesiums.
The anisotropic nature of $\rho _{i}({\bf r})$ around the
Mg ions can be understood in tight-binding terms as arising from
the mixing of Mg(3s) and Mg(3p) states because of the electric
field at the surface. This mixing will also cause a lowering of
one of the 3s/3p hybrid states, and this is one way of
understanding the appearance of gap states below the conduction
band minimum (cf. discussion in references \cite{k11,k10}).

We have also examined the valence electron density in the
surface region, but we do not show it here, since the surface
distribution will become clear in the next Section.

\subsection{The surface F-center}

We have investigated the properties of surface F-centers using
calculations on 2-layer, 4-layer and 6-layer slabs. The same
surface unit cell was used in all cases. The translation vectors
joining nearest neighbor F-centers are in the (110) direction,
and their length is $2\sqrt{2}d$, where $d$ is the cation-anion
separation in the perfect crystal. With this surface unit cell,
the repeating unit contains 16, 32 and 48 sites in the three
slabs. For the 4-layer slab, we have studied F-centers both in
the surface layer and in the layer immediately below the surface,
and for the 6-layer slab, we have also studied the F-center in
the third layer. Each repeating unit contains only a single F-center,
 so that the slab contains an F-center only at one of the
surfaces. The vacuum width and the ${\bf k}$-point sampling are the same
as for the calculations on the perfect surface. In all cases we
have investigated both the unrelaxed and the fully relaxed
systems.

We report in Table 2 two oxygen removal energies for each
case. The first removal energy refers to the energy required to
extract an oxygen atom from the slab with all atoms held fixed at
their perfect lattice positions. The second removal energy refers
to the same process except that both the initial and the final
systems are fully relaxed. It is the second of these energies
that is relevant to the real physical process. Four important
conclusions emerge from these results. The first is that
relaxation effects are extremely small. The largest difference
between the two removal energies is only 0.06 eV, which is
insignificant for practical purposes. The second conclusion is
that the removal energies are insensitive to the slab thickness.
Even for the 2-layer slab the removal energy for the top layer
differs by only 0.2 eV from the value for the 6-layer slab. The
differences between the 4-layer and 6-layer slabs are even
smaller. This indicates that the results are well converged with
respect to the thickness of the slab. The third conclusion is
that the removal energies for the second and third layers are
extremely close to the bulk value of 10.55 eV reported in Table 1.
However, the first-layer removal energy of 9.77 eV is
substantially lower than the bulk value. This effect is expected
became of the 5-fold coordination of the surface oxygen atoms.

As in the case of the bulk F-center, the ionic relaxations
are very small. For the third-layer F-center, the relaxations are
essentially the same as for the bulk F-center. For the surface F-center,
 the neighboring Mg ions relax away from the F-center in
the plane of the surface by 0.02 \AA, which is about twice the
relaxation found in the bulk. In addition, they have a component
of displacement along the outward surface normal which is 0.04 \AA\
with respect to their relaxed surface positions. This implies a
net outward relaxation of 0.02 \AA\ of these neighboring Mg ions
with respect to their positions on the perfect surface. The Mg
ion directly below the first-layer F-center relaxes away from the F-center
 by 0.05 \AA. The relaxation of neighboring oxygens is
negligible.

The electronic DOS for the systems containing first-layer
and third-layer F-centers are shown on the bottom and middle
panels in Fig. 5. The DOS for the
second-layer F-center is almost the same as for the third-layer
case. The electronic level of the unrelaxed first-layer F-center
is 2.0 eV above the top of the valence band, which becomes
2.3 eV on allowing the system to relax, so that relaxation
effects are significant. This F-center energy is slightly below
that of the bulk F-center, which is 2.7 eV above the valence band
maximum (see Section 3).
So far as we are aware, there is no direct experimental evidence
on this question. However, it may be relevant that
a downward shift of about
0.2 eV in the luminescence of surface F-centers (with respect
to that of bulk F-centers) was observed for MgO in
experiments on absorboluminescence \cite{ds13}.
A slight downward shift of about 0.04 eV in the surface F-center
absorption band in KCl crystal which is similar to MgO crystal studied here,
was also reported in \cite{ds12}.
The shift of the F-center level in the direction of lower energies
found in our calculations can be
explained by a competition of a number of factors which work
in opposite directions.
Indeed, it was already mentioned above that the total Coulomb (Hartree)
potential becomes weaker as one moves from the bulk  to the surface.
At the same time, the F-center wave function is more diffuse in this
case as can be judged from our F-center electronic density distribution
which is discussed below.
Both these factors work in lifting up the F-center level. On the other
hand, the positive kinetic energy of a more diffuse F-center state becomes
smaller which results finally in a small downward shift of the F-center level.
For the second layer F-center the
unrelaxed and relaxed values for the F-center level above the VBM
are 2.5 and 2.56 eV, whereas for the
third-layer F-center the value is 2.6 eV, which is very close to
the bulk value.
 It is worth commenting briefly that the band gap for the
first-layer F-center system is slightly greater than for the perfect
surface (3.5 eV rather than 3.2 eV). Such an effect cannot occur for an
isolated F-center, and must be due to the finite concentration of
F-centers because of the periodic boundary conditions.

The valence electron density of the system containing the
first-layer F-center is shown in Fig. 7. The electronic density
shows a maximum near the center of the vacancy, and it is
interesting to note that this maximum value (0.13 electrons \AA$^{-3}$) is
only slightly below the corresponding maximum of the bulk F-center
(0.15  electrons \AA$^{-3}$). The electronic density decays rapidly along the
outward surface normal. The conclusion is that the F-center is
highly localized in this 5-fold coordinated environment. We have
also examined the electron density of the surface F-center level
itself, and we found that it is extremely similar to that of the
bulk F-center. The valence electron distributions for the systems
containing second and third layer F-centers show no significant
differences from the distributions for the bulk F-center.
Our results show that already the third-layer F-center
is very similar to the bulk F-center.

\section{F-center at surface irregularities}

\subsection{The perfect step, corner and reverse corner}

We have studied three kinds of surface irregularities: the
step, the corner and the reverse corner, as shown in Fig. 8. As
before, the calculations are done using slab geometry, but the
repeating unit is now somewhat more complicated. For the step
calculations, we use a repeating cell containing 52 sites with
each surface of the slab consisting of terraces of width $3d$,
separated by steps of height $d$. The slab contains four layers
of ions and the vacuum width is $3d$. We expect this thickness of
the slab to be adequate because the calculations described in the
previous Section show that the oxygen removal energy from the
flat surface is essentially the same for the 4-layer and 6-layer
slabs. In order to achieve this geometry for the step system,
the translation vectors
${\bf A}_{i}$ of the supercell are
$$
{\bf A}_{1}= d(3,2,1), {\bf A}_{2}= d(0,4,0), {\bf A}_{3}= d(0,0,7).
\eqno (4)
$$
The  orientation of the coordinate system corresponding to
this choice of the lattice vectors
is clear from Fig. 8. (It will be noted that for the perfect
step, the vector ${\bf A}_{1}$ could be equally well be taken to be
$d(3,0,1)$; the reason for our choice of ${\bf A}_{1}$ will be explained
below.) The corner and reverse corner systems are studied using
a repeating cell of 44 sites. Each surface of the slab
consists again of terraces and steps, but now the steps zig-zag
back and forth along the (100) and (010) directions. Corners
and reverse corners alternate, with the distance between each
corner and the neighboring reverse corner being $2d$. The slab
thickness and the vacuum width are the same as before. The
translation vectors of the supercell are
$$
{\bf A}_{1}= d(2,-2,0), {\bf A}_{2}= d(2,3,1), {\bf A}_{3}= d(0,0,7).
\eqno (5)
$$
The lowest order Monkhorst-Pack set of two ${\bf k}$-points is used.

The relaxation energies of both the step and the corner
systems are substantial. For the step system, the total
relaxation energy per repeating cell is 1.51 eV, which
corresponds to 0.38 eV per ion pair along the length of the step.
The total relaxation energy for the corner system is 3.01 eV per
cell, or 0.75 eV per ion pair along the length of the zig-zag
step.

The ionic displacements away from their perfect-lattice
positions for the four inequivalent atoms at the step itself are
reported in Table 3. Our
results show that the displacements are quite substantial and
their general direction is such as to smooth out the
discontinuity at the step.
The ionic displacements associated with the corner and the
reverse corner system are reported in Table 4.
Once again, we see substantial displacements having
the same general tendency as for the perfect step. It is worth
noting that the displacements of the Mg ion number 4 at the
bottom of the reverse corner and of its oxygen neighbors 5 are
very small. The step on MgO (001) surface has
previously been studied both by empirical modeling \cite{k9,k4,k6} and by
semi-empirical quantum calculations \cite{k1,ds2}, but we cannot provide
a detailed comparison since the displacements have not been reported
in these  earlier studies. However, the magnitudes and directions
of the displacements that can be inferred from the pictorial data
given in \cite{k9,k4,k6,k1,ds2} seem to be in general agreement
with those we have found.

The electronic DOS for the relaxed step is almost identical to that
of the flat (001) surface, and we do not show it here. Just as for
the flat surface, there is a band of surface states pulled down from
the bottom of the conduction band. The band gap for the step
system is 3.3 eV, which is slightly larger than that for the flat
surface (3.2 eV, see Section 4.1), but we do not regard this as significant.
Thus, our results on the band gap for the step system do not support
experimental suggestion \cite{k5} of a substantial
0.85 eV lowering of the excitonic surface states associated with
four-fold coordinated oxygens on the surface with respect to those for
the flat (001) surface. The only other noticeable
effect is a downward shift of the O(2{\it p}) valence
band with respect to the O(2{\it s})
band by about 0.2 eV.

The DOS of the corner - reverse corner system
shown in Fig. 9 is much more interesting. First, there is a narrow
band of states completely split off from the top of the O(2{\it s}) states
by ca. 1.0 eV.
Second,
there is a significant peak at the top of the O(2{\it p}) valence band,
though this is not split off. Third, there is a band of states split off from
the bottom of the conduction band. The gap between the top of the
valence band and the lowest of the unoccupied states split off
from the bottom of the conduction band is 2.4 eV, which is
considerably less than that  for the flat surface and the
step. The UV diffuse reflectance spectroscopy
\cite{k5} has also demonstrated almost 2.0 eV lowering of the surface
excitonic states attributed to the three-fold coordinated surface oxygen
atoms (i.e. at the corners) with respect to the band gap for the flat (001)
surface. Previous cluster
LMTO calculations \cite{k10}
which found a substantial reduction of the gap at the corners compared
with the flat surface of a cluster are also
consistent with our calculations.
We think that the appearance
of the surface states above the bands of O(2{\it s}) and O(2{\it p}) states
can be simply explained by a drastic decrease of the Hartree
potential at the corner site (as follows from \cite{madsurf}
(see also \cite{rew}, p. 117), the Madelung
constant at the corner site is only 0.87378, i.e. half the value
 for the five-coordinated surface site where it is 1.68155).
The fact that
the surface states
are not split off completely from the top of the upper valence
band is the result of a more delocalized character of these states
in comparison with the O(2{\it s}) surface states.

As in the case of the flat surface, we have investigated the nature of the
surface states
by calculating the electron density associated with them, see Eq.(3).
For the step system, we find that the surface states at the bottom of
the conduction band are extremely similar to those at the flat
surface.  For the corner - reverse corner system, the surface states
split off from the O(2{\it s}) band have {\it s} character and are
strongly localized on the corner oxygens with a substantial contribution
from the three nearest oxygens on the terrace underneath.
The surface states at
the top of the valence band have {\it p} character and are also
localized mainly on the corner oxygens; the axis of the {\it p}-like
density distribution passes through the corner site and lies almost
in the (111) direction.  The states at the bottom of the conduction
band are shown in Fig. 10, which indicates that they are strongly
associated with Mg ions neighboring the reverse corner, though there
is also considerable weight on surface oxygen ions.

The total spatial densities of valence electrons in the
perfect step and corner systems do not show very remarkable
features. However, in order to provide a point of reference for
the F-center densities shown later, we display in Fig. 11 the
density associated with the corner system on a plane parallel to
the terraces passing through the zig-zag step. It is noteworthy
that the large displacement of the corner oxygens leads to a
significant build-up of charge between these ions and their
oxygen neighbors.

\subsection{F-center at the step, corner and reverse corner}

We have calculated the energy for removing oxygen from the
step, the corner and the reverse corner. In the case of the step,
we have positioned the F-centers so as to maximize the separation
between them, as shown in Fig. 8{\it a}. This is the reason for taking
the translation vectors ${\bf A}_{i}$ for the step as shown in Section
5.1 (see Eq.(4)).
In Fig. 8{\it b}, we show the location of the F-centers at the
corner and the reverse corner.

The unrelaxed and relaxed oxygen removal energies are given
in Table 5. We note the following points. At the reverse corner,
the oxygen removal energy is only slightly less (by 0.4 eV) than
its value for the first-layer F-center on the flat surface. In
both these cases, the F-centers are 5-fold coordinated. A further
reduction of 0.4 eV occurs when the F-center is formed at the 4-fold
coordinated step site, and the removal energy decreases by
0.9 eV more when we go to the 3-fold coordinated corner site.
Overall, the removal energy is 2.5 eV lower at the corner site
than in the bulk. The difference between unrelaxed and relaxed
removal energies is very small except for the corner site, where
it is a little over 1 eV.

The ionic displacements around the F-center for the step system
are reported in Table 3, whereas those for systems with a F-center
at the corner and reverse corner sites are given in Table 4. Although
the general trend is the same as in all systems discussed so far, namely
it is an additional outward movement of nearest Mg ions from the
F-center site, the additional displacements are found to be substantial only
for the F-center at the step and at the corner.

The electron DOS for the system containing  the F-center at the step
is shown in Fig. 12. This is essentially identical to the DOS for the
perfect step system, except for the presence of the narrow F-center
band in the gap. The mean energy of this band lies 2.4 eV above the top
of the valence band, which is almost identical to its position
for the F-center on the flat surface. The DOS for the F-center
at the corner and the reverse corner are shown in Fig. 9. The
simplest case to understand is that of the F-center at the reverse
corner, for which the DOS is very similar to that of the perfect
corner - reverse corner system except for the presence of the
F-center band in the gap, whose mean position lies only 0.8 eV above
the top of the band of O(2{\it p}) surface states.  The DOS of the
corner F-center is more complicated, because in addition to the
appearance of the F-center band, there also are substantial changes
to the surface states. The mean position of the F-center band is 1.7
eV above the top of the band of O(2{\it p}) surface states, which is
considerably higher than that for the reverse corner case. This can mainly be
explained by the reduction of the Hartree potential at the corner site
and by the strong delocalization of the F-center wave function
associated with the F-center band.

We have also looked at the spatial distribution of some interesting
features in the DOS. We found that because of the absence of oxygen atoms
at the corner sites on the upper side of the slab,
the states associated with the band split off from the top
of the O(2{\it s}) band have changed their character of localization.
Namely, they appear to be more delocalized over a larger number of oxygens
which are behind the ``empty'' corners on the same terrace.
 The localization
of these states in the two lowest layers is not changed. Surprisingly,
the energy position of this band is also not changed with respect to
that for the perfect corner - reverse corner system.
In addition to that, we have found similar changes in the nature of
states associated with
the surface states at the top of the O(2{\it p}) band. Unoccupied states
attributed to the band split off from the bottom of the conduction
band also show considerable delocalization over the corner regions
along the direction of the zig-zag, as well as
over oxygen atoms which are behind the corner sites on the same
terraces.

The electron densities associated with the F-center state at the step and the
reverse corner are qualitatively rather similar to what we have
already shown for the flat (001) surface. As in that case, the density
is strongly localized at the F-center site, with substantial weights on
neighboring oxygens. The height of the peak in the density at the
reverse corner F-center site is almost identical to that found at the
surface F-center, as might be expected since both sites have the same
coordination.  The height of the peak at the step F-center site is
about 30 percent lower.  On the other hand, the density associated
with the corner F-center state shows considerable delocalization as
demonstrated in Fig. 13.

\section{Discussion}

We first comment on the relaxed structure and the electronic
structure of the MgO surface without F-centers. Our results
for the flat (001) surface confirm what is already known from LEED
\cite{s6,s2,welton}
measurements, from previous {\em ab initio} \cite{s5} and
semiempirical
quantum calculations \cite{sjacek}, and from modelling based on interaction
models \cite{k9,s8} that relaxation effects are extremely small, and consist
mainly of rumpling, with O ions moving slightly out of the surface
and Mg ions moving slightly in. However, we have shown that
relaxation effects are far larger at the step, corner and
reverse corner, and are associated with large energy reductions. The
ionic displacements are such as to smooth out the surface
irregularity. This is easy to understand since the (001) surface of MgO
crystal is the most stable one \cite{rew,k6,sjacek}. Our displacements
agree qualitatively with the displacements
obtained in the model-based calculations \cite{k9,k4,k6},
but in the absence of published
quantitative results from the latter work we are unable to make
detailed comparisons.

The most important new information from this
part of our work concerns the surface electronic states.  We have
shown that there is a substantial narrowing of the band-gap at the
flat (001) surface by ca. 1.6 eV, and that this is entirely due to the
formation of surface states at the bottom of the conduction band. We
find no surface states above the top of the valence band, and nor do
we find any surface states associated with the O(2{\it s}) band.  These
findings are in accord with recent LMTO LDA work \cite{s13}
where detailed calculations
of the band structure of the (001) MgO surface have been done.
Surface states pulled down from the  CBM have been found there,
the bottom of this band
lying ca. 0.6 eV below the bottom of the bulk conduction band.
Our results on the predominant localization on surface oxygens of
the surface states associated with the bottom of the conduction
band support recent theoretical semiempirical calculations \cite{shluexc}.
Namely, it was shown that surface excitonic transitions in MgO
are of the one-center-type, i.e. are mainly localised on
surface oxygen atoms.

According to our calculations, the DOS of the perfect step is
essentially the same as for the flat surface, but important new features
appear for the corner - reverse corner system. First, a prominent band of
surface states splits off from the O(2{\it s}) band. The existence of
this type of surface state is well known from {\em ab initio} calculations
on other oxides such as TiO$_{2}$ \cite{2s-split}.
In the present case,
it seems to exist only at corners, and this suggests that it could be
used as an experimental tool for detecting the presence of corners.
The second new feature is the appearance of surface states above the
bulk VBM, which leads to a further narrowing of the surface band-gap by
ca. 0.8 eV below its value for the flat surface. This might also be used as
an experimental tool for detecting corners.

The systematic reduction of surface band-gap as one passes from the bulk
to the flat (001) surface and then to the corner irregularity may already
have been observed in  the UV diffuse reflectance
spectroscopy \cite{k5}, which indicates
a series of peaks below the fundamental
absorption edge (which was found to be at 8.7 eV). First
two peaks
at 4.62 eV and 5.75 eV   have been
attributed to bound (localised)
excitonic transitions near 3-fold (corners) and
4-fold (steps) surface oxygens. Another two peaks at 6.6 eV and 7.7 eV
have been associated with free excitons on the flat surface and in the
bulk, respectively.
These experimental observations are generally consistent with
what we have obtained in our calculations on the flat surface
and various surface irregularities. The same trend for the band gap
was also found in DV-X$_{\alpha}$ cluster calculations \cite{k10}.

Turning now to our results for the F-center, we note first that our
calculated properties of the bulk F-center are in close agreement
with the previous calculations of Wang and Holzwarth \cite{db3}.
Our finding that the bulk F-center level lies 2.7 eV above the VBM
is very similar to their energy of 2.3 eV, and our conclusions about
the strong localization of the total electron density at the F-center
site but the much weaker localization of the density associated
with the F-center level are in complete agreement with what they find.
We also agree about the almost total absence of relaxation effects.
The new information we have added is that the results are completely
independent of the size of repeating cell, provided the number of sites
is at least 16. We can therefore regard our results
as referring to an isolated F-center in an infinite crystal.

Our calculated value of 10.5 eV for the bulk oxygen removal energy
can be related to the  F-center
formation energy Q defined as the oxygen removal energy minus
the cohesive energy per formula
unit with respect to free atoms \cite{db3,ds1}.
Using our calculated cohesive energy of 11.1 eV, we obtain the value
$Q = -0.6$  eV. The calculations of Wang and Holzwarth \cite{db3}
gave the value $Q = -0.4$ eV. In principle, $Q$ can be measured by
additive coloring experiments in which the equilibrium between bulk
F-centers and Mg vapour is studied as a function of temperature and
pressure. There has been only one attempt \cite{add} to measure $Q$ for MgO,
which gave the value 1.5 eV. The calculated values obtained both by us and by
Wang and Holzwarth are clearly in poor agreement with this.
However, substantial errors are quoted in the experimental measurements,
and we believe that the measured $Q$ is subject to an error of at least
$\pm 0.8$ eV. In addition, there may well have been systematic errors
due to lack of equilibration between bulk and vapor. We therefore
believe that this disagreement is not necessarily due to inaccuracy
in the calculations.

Our calculations have shown the expected reduction of oxygen
removal energy as the coordination number of the oxygen site
decreases through the sequence: bulk, flat surface, reverse corner,
step, corner. The removal energy actually decreases rather little,
being only 2.5 eV or 20 \% lower at the corner than it is in the
bulk. Other properties of the F-center also seem to depend rather
weakly on its location. As we go from the bulk to the flat surface, the degree
of localization of the F-center electron density (as measured by
the value of the density at the F-center site) scarcely changes, the
energy of the F-center level decreases by only 0.4 eV, and
relaxation effects continue to be negligible. The further changes in
electron distribution and the energy of the F-center level on going to the
step site are almost negligible. It is only on going to the corner
and reverse corner sites that significant changes occur. Namely,
the F-center level is found surprisingly close to the top of the
valence band for the 5-fold coordinated reverse corner system
(0.8 eV). However, as follows from our calculations, for the 3-fold
coordinated corner
system, for which the F-center electronic density is substantially
delocalised, its level is further pulled up by 0.9 eV.

\section*{Acknowledgments}

The work of LNK and JMH is funded by EPSRC grants GR/J37546 and
GR/H67935.  The work of LNK was also supported in part by the Latvian
Scientific Council, grant 93.270.
The calculations were performed using an allocation of
time on the Cray T3D at EPCC provided by the High Performance
Computing Initiative.  For assistance in enhancing the speed of the
CETEP code we are indebted to Dr. I. Bush of Daresbury Laboratory.
Analysis of the results was performed using local distributed
hardware funded by EPSRC grants GR/H31783 and GR/J36266. We
gratefully acknowledge useful
discussions with Prof. R. Joyner, Prof. A. M. Stoneham
FRS, Dr. A. H. Harker, Dr. A. L. Shluger, Prof. Dr. J.-M.Spaeth and
Dr. R. Williams.

\newpage
{\bf Table 1.} Calculated oxygen removal energies (in eV) for both
unrelaxed and relaxed F-center systems of various sizes of the repeating
cells in the bulk of MgO crystal.
\vskip 1cm
\begin{center}
\begin{tabular}{||c|c|c||}
\hline
number of atoms &  & \\
in the cell     & unrelaxed & relaxed \\
\hline
8  & 10.656 & - \\
16 & 10.554 & 10.553  \\
32 & 10.579 & 10.571  \\
54 & 10.568 & 10.560   \\
64 & 10.568 & 10.547  \\
\hline
\end{tabular}
\end{center}


\vskip 2cm
{\bf Table 2.} Oxygen removal energies (in eV) for both
unrelaxed and relaxed surface F-center in various positions
with respect to the surface using calculations on
2-layer, 4-layer and 6-layer slabs.
\vskip 1cm
\begin{center}
\begin{tabular}{||c|c|c|c|c|c|c||}
\hline
number of layers & \multicolumn{6}{c ||}{The F-center position} \\
\cline{2-7}
in the slab & \multicolumn{2}{c |}{1st layer}
            &   \multicolumn{2}{c |}{2nd layer}
            & \multicolumn{2}{c ||}{3rd layer} \\
\cline{2-7}
    & unrelaxed & relaxed & unrelaxed & relaxed & unrelaxed & relaxed \\
\hline
2   & 9.618  & 9.556 & - & - & - & - \\
4   & 9.801  & 9.759 & 10.396  & 10.397 & - & - \\
6   & 9.811  & 9.768 & 10.416  & 10.414 & 10.524 & 10.514  \\
\hline
\end{tabular}
\end{center}
\newpage

{\bf Table 3.} Displacements of ions shown in Fig. 8{\it a} at the perfect
step and around the F-center at the step given in units of the
nearest neighbors O - Mg distance in the perfect lattice
$d = 2.082$ \AA.

\vskip 1cm
\begin{center}
\begin{tabular}{||c|c|c|c|c|c|c||}
\hline
Ion & Type &\multicolumn{2}{c |}{Perfect step}
            &   \multicolumn{3}{c ||}{F-center system} \\
\cline{3-7}
  &  & x & z & x & y & z \\
\hline
1 & Mg &  0.12 & 0.02  & 0.11  & -0.07 & 0.06 \\
2 & O  &  0.09 & 0.05  &  -    &  -    &  -    \\
3 & Mg &  0.04 & -0.01 & 0.07  & 0.0   & 0.01 \\
4 & O  &  0.05 &  0.0  & 0.05  & 0.01  & 0.0 \\
5 & O  & -0.03 &  0.06 & -0.02 & 0.01  & 0.07      \\
6 & Mg & -0.04 &  0.07 & -0.03 & 0.0   & 0.06     \\
\hline
\end{tabular}
\end{center}
\vskip 2cm

{\bf Table 4.} Displacements of ions shown in Fig. 8{\it b} at the perfect
corner - reverse corner system as well as for the same systems
with F-center at the corner and the reverse corner positions,
given in units of the
nearest neighbors O - Mg distance in the perfect lattice
$d = 2.082$ \AA.

\vskip 1cm
\begin{center}
\begin{tabular}{||c|c||c|c|c||c|c|c||c|c|c||}
\hline
 &  & \multicolumn{3}{c ||}{}
            & \multicolumn{6}{c ||}{F-center at } \\
\cline{6-11}
Ion & Type & \multicolumn{3}{c ||}{Perfect system}
            & \multicolumn{3}{c ||}{the corner}
            & \multicolumn{3}{c ||}{the reverse corner} \\
\cline{3-11}
  &  & x & y & z & x & y & z & x & y & z \\
\hline
1 & O  & -0.04 & -0.04 & 0.03 & 0.02 & 0.02 & 0.01 & -     &  -    &  -  \\
2 & Mg &  0.01 &  0.11 & 0.00 & 0.07 & 0.07 & 0.05 & 0.0   & 0.08  & 0.04 \\
3 & O  &  0.11 &  0.11 & 0.06 & -    &  -   &   -  & 0.10  & 0.10  & 0.05 \\
4 & Mg & -0.01 & -0.01 & 0.00 & 0.0  & 0.0  & 0.01 & -0.01 & -0.01 & -0.04 \\
5 & O  &  0.00 & -0.02 & 0.02 & -0.01& -0.01& 0.04 & 0.02  & -0.01 & 0.02 \\
6 & Mg & -0.04 & -0.04 & 0.13 & -0.02& -0.02& 0.01 & -0.02 & -0.02 & 0.13 \\
7 & Mg & 0.01  &  0.02 & -0.02& 0.02 & 0.02 & -0.01& 0.02  & 0.07  & -0.01 \\
8 & O  &  0.06 &  0.06 & 0.00 & 0.03 & 0.03 & 0.01 & 0.06  & 0.06 & 0.01 \\
\hline
\end{tabular}
\end{center}
\newpage

{\bf Table 5.} Oxygen removal energies (in eV) for
unrelaxed and relaxed F-center at the step, the corner
and the reverse corner systems.

\vskip 1cm
\begin{center}
\begin{tabular}{||c|c|c|c|c||}
\hline
type of      &  &  &  & \\
irregularity & nn$^\dagger$ & nnn$^\ddagger$ & unrelaxed & relaxed   \\
\hline
step   & 4 & 5 & 8.931  & 8.997  \\
corner & 3 & 3 & 7.012  & 8.058  \\
reverse corner & 5 & 7 & 9.205  & 9.367  \\
\hline
\end{tabular}
\end{center}
$^\dagger$ Number of nearest Mg ions around the vacancy

\noindent $^\ddagger$ Number of nearest oxygen ions around the vacancy
\newpage

\centerline{\bf Figure captions}
Fig. 1. Electronic DOS of the perfect bulk crystal (top panel) and
the 16-site (middle panel) and 54-site (bottom panel) bulk F-center systems.
Arbitrary units are used for the DOS.

Fig. 2. A contour plot of the valence electronic density of the 54-site
F-center system (in units of $10^{-2}$ electron/\AA$^3$).
The cut has been made
in the (001) plane. To avoid high peaks on oxygens,
the density has been chopped at 0.2 electron/\AA$^3$. The positions
of oxygen atoms are indicated by the symbol X. Distances are in \AA.

Fig. 3. The dependence of $\Delta n(R)$ (see Eq.(2)) for the 54-site
F-center system on the sphere radius $R$ (see text for details).

Fig. 4. A contour plot of the electronic density associated with
the F-center band for the 54-site F-center system. Positions of
Mg atoms are indicated by the symbol $\Delta$. The units and other
notations are the same as in Fig. 2.

Fig. 5. The total DOS of the unrelaxed six-layer slab (top panel) and
the flat surface systems containing third layer (middle panel)
and  first layer (bottom panel) F-centers.
Arbitrary units are used for the DOS.

Fig. 6. A contour plot of the fictitious charge density (see Eq. (3))
associated with the first unoccupied state at the bottom of the
conduction band of the flat  surface. The cut has been made parallel
to the surface normal through nearest surface Mg and O ions. The top two
layers of atoms are shown explicitly.
The units and notations are as in Fig. 4.

Fig. 7. The contour plot of the valence electronic density of the
flat surface system containing the first-layer F-center. The cut has been
made along the direction of the surface normal through the F-center and
nearest Mg ions on the surface. Other notations and units
are the same as in Fig. 2.

Fig. 8. Repeating geometries used in the {\it ab initio} calculations
for the step ({\it a}) and the corner - reverse corner ({\it b}) systems.
Mg and O atoms are indicated by grey and white circles, respectively.
In each case, only upper side of the slab is shown. Note that
the axes of the Cartesian coordinate system are also shown for each
case. They correspond to Eqs.(4) and (5). Corresponding F-center systems
have been created by removing oxygen atoms 2 ({\it a}), 3
or 1 ({\it b}) in the case of the step, corner and reverse corner
systems, respectively.

Fig. 9. The electronic DOS of the perfect corner - reverse corner
system (top panel), the system containing the F-center at
the corner (middle panel) and the reverse corner (bottom panel).
Arbitrary units are used for the DOS.

Fig. 10. A contour plot of the fictitious charge density (see Eq. (3))
associated with the band of unoccupied states split off from the
the bottom of the
conduction band of the perfect corner - reverse corner system.
The cut has been made in a plane
parallel to the terraces passing through the corners along the zig-zag
step.  Other notations and units are the same as in Fig. 4.

Fig. 11. A contour plot of the
total valence electronic density for the perfect
corner - reverse corner system.
The cut is the same as in  Fig. 10.
Other notations and units are the same as in Fig. 2.

Fig. 12. The electronic DOS for the system containing the F-center at the
step.
Arbitrary units are used for the DOS.

Fig. 13.  A contour plot of the electronic density associated with
the F-center band for the corner F-center system.
The cut is the same as in  Fig. 10.
Other notations and units are the same as in Fig. 4.

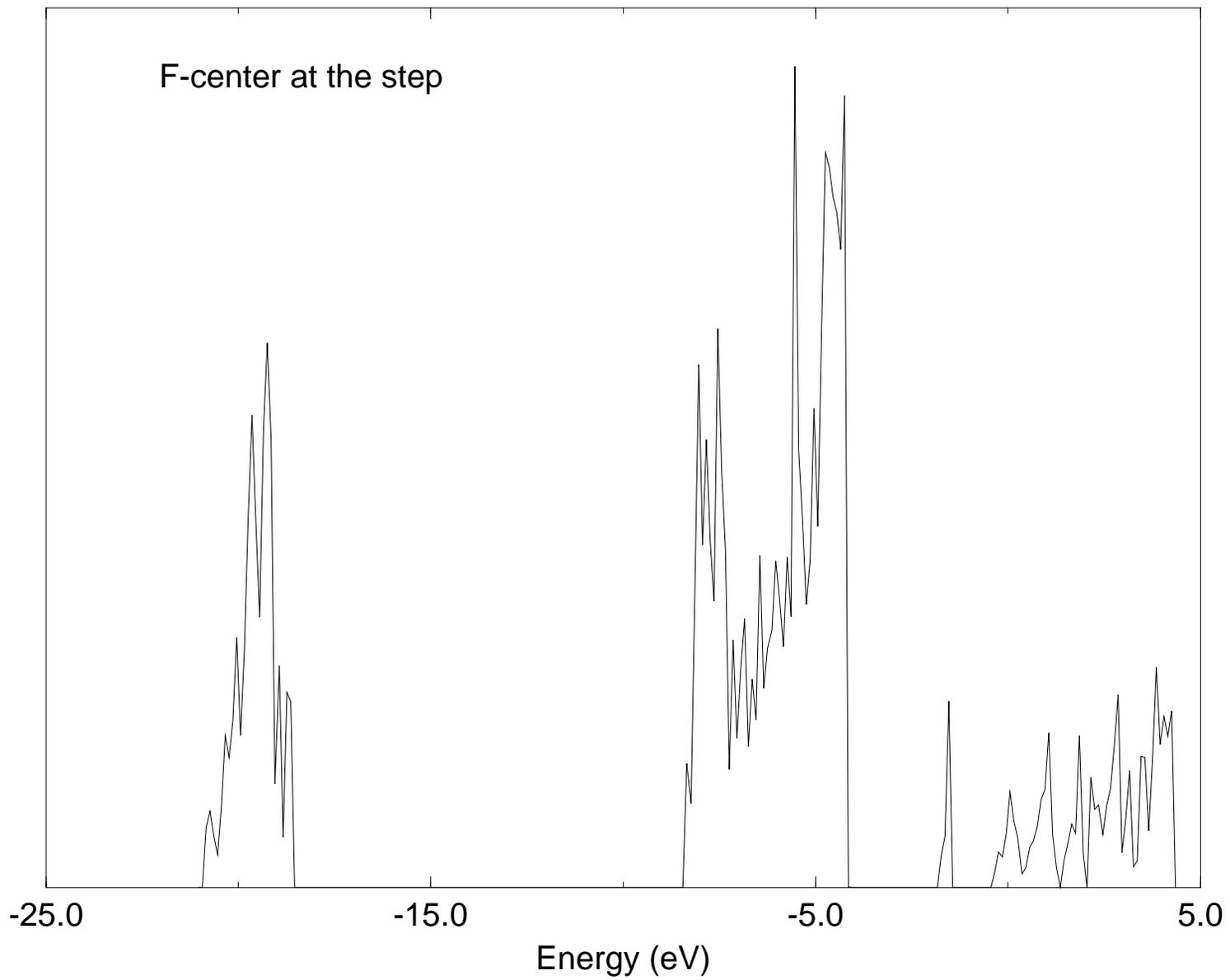
\end{document}